\documentclass[aps,pra,twocolumn,longbibliography,groupedaddress,showpacs,floatfix]{revtex4-1}
\usepackage{mathtools,amssymb,graphicx,units,xfrac,bm,bbold}
\usepackage{lmodern}
\usepackage[plainpages=false,
	pdfpagelabels,
	colorlinks=true,
	linkcolor=red,
	urlcolor=blue,
	citecolor=blue,
	pdftitle={Title},
	pdfauthor={},
	pdfdisplaydoctitle=true,
	pdfduplex=DuplexFlipLongEdge]{hyperref}
\usepackage{soul} 
\usepackage[dvipsnames,usenames,svgnames]{xcolor}
\usepackage[normalem]{ulem}
\usepackage{color}
\graphicspath{{./Figures/}} 

\usepackage{lipsum}

\emergencystretch=\maxdimen
\DeclareMathOperator{\sign}{sign}

\def\rm#1{\mathrm{#1}}
\def\bf#1{\mathbf{#1}}
\def\rb#1{\mathrm{\mathbf{#1}}} 
\def\ave#1{\langle\, #1\,\rangle}

\begin{document}
\normalem

\title{
Magnetism and metal-insulator transitions in the anisotropic kagome lattice
}
\author{Lucas O.~Lima}
\author{Andressa R.~Medeiros-Silva}
\author{Raimundo R.~dos Santos}
\author{Thereza Paiva}
\author{Natanael C.~Costa}
\affiliation{Instituto de F\'isica, Universidade Federal do Rio de Janeiro, Rio de Janeiro, Rio de Janeiro 21941-972 , Brazil}
%
%
\begin{abstract}
The interest in the physical properties of kagome lattices has risen considerably. In addition to the synthesis of new materials, the possibility of realizing ultracold atoms on an optical kagome lattice (KL) raises interesting issues.  For instance, by considering the Hubbard model on an anisotropic KL,  
with a hopping $t^\prime$ along one of the directions, one is able to interpolate between the Lieb lattice  ($t^\prime=0$) and the isotropic KL ($t^\prime=t$). The ground state of the former is a ferrimagnetic insulator for any on-site repulsion, $U$, while the latter displays a transition between a paramagnetic metal and a Mott insulator. 
One may thus consider $t^\prime$ as a parameter controlling the degree of magnetic frustration in the system.
By means of extensive quantum Monte Carlo simulations, we have examined magnetic and transport properties as $t^\prime$ varies between these limits in order to set up a phase diagram in the $(U/t, t^\prime/t)$ parameter space.
As an auxiliary response, analysis of the average sign of the fermionic determinant provides consistent predictions for critical points in the phase diagram. We observe a metal-insulator transition occurring at some critical point $U_c^\rm{M}(t^\prime)$, which increases monotonically with $ t^\prime $, from the unfrustrated lattice limit. 
In addition, we have found that the boundary between the ferrimagnetic insulator and the Mott insulator rises sharply with $t^\prime$.
\end{abstract}
\date{\today}

\maketitle

\section{Introduction}
\label{introd}

The interplay between geometrically-induced frustration and fermion itinerancy gives rise to fascinating magnetic states such as quantum spin liquids (QSL's), characterized by highly degenerate ground states \cite{Diep05}. Examples of exotic magnetic phases have become more abundant over the last decades, such as those found in organic materials with triangular lattice structures \cite{Shimizu03,Kagawa04,Lefebvre20}, and in herbertsmithite compounds with a kagome lattice (KL) structure\,\cite{Helton07,Mendels07,Helton10}, although for the latter the nature of the QSL state is still under debate~\cite{Han12,Mingxuan15,Norman16}. More recently, the emergence and continuing development of experiments on optical lattices, in which ultracold atoms are loaded and the interaction amongst them controlled through Feshbach resonances, has enabled the study of strongly correlated systems in an unprecedented controllable way  \cite{Chin10,Jaksch05,Bloch08,Esslinger10,McKay11}. In particular, optical lattices with the kagome structure have been recently realized with bosonic atoms \cite{Jo12,Barter20,Leung20}, and one expects that optical KL's with fermionic atoms will become available in the near future.

\begin{figure}[t] 
	\begin{center}
	\includegraphics[scale=0.75]{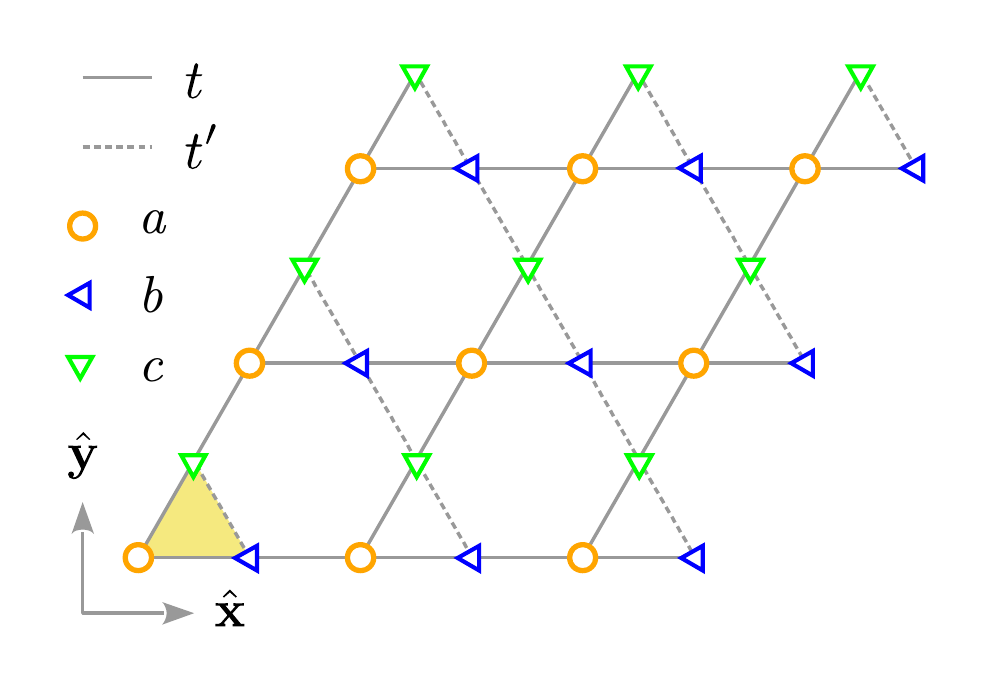}
	\caption{\label{fig:lattice} The kagome lattice is composed of hexagons tiled with intervening corner-sharing triangles; points $a$, $b$ and $c$ comprise the unit cell (highlighted triangle). Solid and dashed lines indicate hopping amplitudes $t$ and $t^\prime$, respectively.}
	\end{center}
\end{figure}

In Fig.\,\ref{fig:lattice} we visualize the KL as a tilted square Bravais lattice with a three-site basis, whose sites are denoted by $\alpha=a, b,$ and $c$. Frustration in the elementary triangles is evident if one allows fermions to hop between sites $\alpha$ and $\alpha'\neq\alpha$, with amplitude $t$, subject to an on-site repulsion, $U>0$, thus giving rise \cite{Hubbard63} to an effective antiferromagnetic exchange interaction, $J\sim t^2/U$, in the strong coupling regime. The kagome lattice additionally displays a tight-binding feature absent in the triangular lattice, namely the presence of a flat band at its high-energy edge; see Fig.\,\ref{fig:dos-bandwidth}(d). Flat bands may lead to a wide range of unexpected electronic properties, such as unconventional magnetism \cite{Yin19,Li21} and superconductivity~\cite{Cao18}. In particular, the Hubbard model on a KL has been recently studied \cite{andressa22} through determinant Quantum Monte Carlo (DQMC) simulations: at half filling it was found that a phase transition between a paramagnetic metal and a Mott insulator occurs at $U_c/t\approx 6.5$.

Figure \ref{fig:lattice} also highlights the fact that if the hopping, $t^\prime$, between $b$ and $c$ sites (i.e., along the dashed lines) is switched off, one ends up with the so-called Lieb lattice (or decorated square lattice, or CuO$_2$ lattice). This lattice also displays a flat tight-binding band, but now located at the particle-hole symmetry (PHS) point; see Fig.\,\ref{fig:dos-bandwidth}(a). With on-site repulsion, the Lieb lattice (LL) at half filling is not frustrated, but is a ferrimagnetic insulator for all $U>0$  \cite{Lieb89,Lieb89err,Costa16}.

A question immediately arising is how the ferrimagnetic insulator at $t^\prime=0$ evolves to either a Mott insulator or a paramagnetic metal as $t^\prime\to t$. So far, studies of the Hubbard model with anisotropic hopping on the KL have been primarily through mean-field--like approaches \cite{Imai03,Furukawa10,Yamada11}. We therefore feel that a thorough investigation of the ground state phase diagram through unbiased methods is certainly in order. With this in mind, here we perform extensive DQMC simulations on the half-filled Hubbard model on the KL with anisotropic hopping, in order to propose a ground state phase diagram, $(U/t, t^\prime/t)$.

The layout of the paper is as follows.
In Sec.\,\ref{sec:Model_Methodology} we present the model and the main features of DQMC method.
In Sec.\,\ref{sec:results}, we highlight the main properties of the noninteracting case, and investigate the magnetic and transport observables of the interacting one.
Sec.\,\ref{sec:conc} summarizes our findings.

\section{Model and Methodology}
\label{sec:Model_Methodology}

In order to interpolate between these two limits (Lieb and Kagome lattices), we write the Hamiltonian as
\begin{equation}\label{eq:Hamiltonian}
	\widehat{\mathcal{H}} = \widehat{\mathcal{H}}_\rm{K} + \widehat{\mathcal{H}}_\rm{U} + \widehat{\mathcal{H}}_\mu,
\end{equation}
where $\widehat{\mathcal{H}}_\rm{K}$ is the kinetic energy, $\widehat{\mathcal{H}}_\rm{U}$ describes the on-site interaction, and $\widehat{\mathcal{H}}_\mu$ controls the band filling.
They are defined as
\begin{subequations}
\begin{align}
\label{eq:Kinetic_energy}
\widehat{\mathcal{H}}_\rm{K} = & -t\, \sum_{\mathbf{r},\sigma}\left( a^{\dagger}_{\mathbf{r},\sigma} b_{\mathbf{r},\sigma} + a^{\dagger}_{\mathbf{r},\sigma} c_{\mathbf{r},\sigma} + \mathrm{H.c} \right) + \nonumber\\
& -t\, \sum_{\mathbf{r},\sigma}\left( a^{\dagger}_{\mathbf{r},\sigma} b_{\mathbf{r} - \hat{\rb{x}},\sigma} + a^{\dagger}_{\mathbf{r},\sigma} c_{\mathbf{r} - \hat{\rb{y}},\sigma} + \mathrm{H.c} \right) + \\
& -t^{\prime}\, \sum_{\mathbf{r},\sigma}\left( b_{\mathbf{r},\sigma}^{\dagger} c_{\mathbf{r},\sigma} + b_{\mathbf{r},\sigma}^{\dagger} c_{\mathbf{r} + (\hat{\rb{x}}-\hat{\rb{y}}),\sigma} + \mathrm{H.c}  \right),  \nonumber\\
\label{eq:Potential_energy}
\widehat{\mathcal{H}}_\rm{U} = & ~ U \sum_{\mathbf{r},\alpha} \left( n^{\alpha}_{\mathbf{r},\uparrow} - \sfrac{1}{2} \right)\left( n^{\alpha}_{\mathbf{r},\downarrow} - \sfrac{1}{2} \right), \\
\label{eq:Chemical_energy}
\widehat{\mathcal{H}}_\mu = & -\mu \sum_{\mathbf{r},\sigma,\alpha} n^{\alpha}_{\mathbf{r},\sigma}  \; ,
\end{align}
\end{subequations}
where $a^{(\dagger)}_{\mathbf{r},\sigma}$, $b^{(\dagger)}_{\mathbf{r},\sigma}$, and $c^{(\dagger)}_{\mathbf{r},\sigma}$ are standard fermion annihilation (creation) operators acting on orbital $\alpha$ at position $\rb{r}$, with spin $ \sigma $; $n^{\alpha}_{\mathbf{r},\sigma}$ are number operators acting on their corresponding orbitals, $\alpha =a$, $b$, or $c$. $ U $ is the strength of the on-site repulsion, and $\mu$ is the chemical potential. The first two terms on the right hand side of Eq.\,\eqref{eq:Kinetic_energy} denote the inter- and intracell hopping between $a$ and $b$-orbitals, or $c$-orbitals, respectively, while the third term corresponds to the inter- and intracell diagonal hopping between $b$ and $c$-orbitals (dashed line in Fig.\,\ref{fig:lattice}), respectively.
Notice that, by varying $t^\prime/t$ between 0 and 1, one continuously increases frustration.
Hereafter, we define the lattice constant as unity, and set the energy scale in units of the hopping integral $t$ of the nearest neighbors $a$-$b/c$ sites.

We examine the properties of Hamiltonian in Eq.\,\eqref{eq:Hamiltonian} by employing the DQMC methodology\,\cite{Blankenbecler81,Hirsch83,Hirsch85,White89,dosSantos03b}, which
provides an unbiased treatment of the interactions in $\widehat{\mathcal{H}}$. 
The idea is based on an auxiliary field decomposition of the interaction that maps the system into free fermions moving in a floating and time-dependent potential (imaginary) space. The method basically consists of two key steps. 
First, in the grand partition function one separates the noncommuting parts of the Hamiltonian through the so-called Trotter-Suzuki decoupling,
\begin{align}\label{eq:TS-decoupling}
\mathcal{Z} &= \rm{Tr}\;e^{-\beta\widehat{\mathcal{H}}} = \rm{Tr}\,[(e^{-\Delta\tau(\widehat{\mathcal{H}}_\rm{K} + \widehat{\mathcal{H}}_{\rm U})})^{M}] \nonumber \\ &\thickapprox \mathrm{Tr}\, [e^{-\Delta\tau\widehat{\mathcal{H}}_\rm{K}}e^{-\Delta\tau\widehat{\mathcal{H}}_{\rm U}}e^{-\Delta\tau\widehat{\mathcal{H}}_\rm{K}}e^{-\Delta\tau\widehat{\mathcal{H}}_{\rm U}} \cdots],
\end{align}
where $M=\beta/\Delta\tau$, with $\Delta\tau$ being the grid of the imaginary-time axis and $ \beta=1/(k_{\rm{B}}T) $ is the inverse temperature, where $ k_{\rm{B}} $ is the Boltzmann constant.  
This decomposition leads to an error proportional to $(\Delta\tau)^{2}$, which can be systematically reduced as $\Delta\tau\to 0$. Here, we choose $\Delta\tau\leq0.1$, which is small enough so that the systematic errors from the Trotter-Suzuki decoupling are comparable to the statistical ones (i.e., from the Monte Carlo sampling).

The second step is to perform a discrete Hubbard-Stratonovich (HS) transformation \cite{Hirsch83} on the two-particle terms, $\exp{(-\Delta\tau\widehat{\mathcal{H}}_{\rm U})}$, which converts them also to quadratic form in fermion operators, at the cost of introducing discrete HS auxiliary fields, $s(\rb{r},\tau)$. 
In this way the resulting trace of fermions propagating in an auxiliary bosonic field
can be performed. 
Thus, one can evaluate Green's functions and other physical observables including spin-, charge- and pair correlation functions by sampling the HS fields with the product of fermionic determinants acting as Boltzmann weights, i.e.,
$p(s)=\rm{det}\,M_\uparrow(s(\rb{r},\tau))\,\rm{det}\,M_\downarrow(s(\rb{r},\tau))$. 
However, these are not positive definite and when averaging an observable, $\mathcal{O}$, one takes $|p(s)|$ at the cost of keeping track of the average sign; that is,
\begin{equation}
\label{eq:sign}
    \ave{\mathcal{O}} = \frac{\sum_s |p(s)|\,\sign(s)\,\mathcal{O}(s)}{\sum_s |p(s)|\,\sign(s)}\equiv \frac{\ave{\sign \times \mathcal{O}}}{\ave{\sign}}.
\end{equation}
 The positiveness of the product of determinants is guaranteed for systems displaying PHS at half-filling, such as bipartite lattices (e.g., square, honeycomb and Lieb) and for attractive interactions. 
 
 Apart from these cases, and depending on band filling, temperature, interaction strength, lattice geometry and size, and so forth, one may end up with $\ave{\text{sign}}\ll 1$,  thus rendering $\ave{\mathcal{O}}$ meaningless: this is the infamous `minus-sign problem' \cite{Loh90,Troyer05,Mondaini12,Iglovikov15}.
 Nonetheless, it has been recently suggested  that low values of $\ave{\text{sign}}$ may be used to locate critical points \cite{Mondaini22,mondaini2022_2}. For the case at hand, we note that the KL is not bipartite, so that PHS is absent for any filling; hence, one must keep track of $\ave{\text{sign}}$; see below.
 
 Through the DQMC algorithm we obtain Green's functions, from which several quantities are calculated; see. e.g.\ Refs.\, \cite{Blankenbecler81,Hirsch85,dosSantos03b}.
 Therefore, the magnetic response of the system may be probed by the real space spin-spin correlation functions,
\begin{equation}\label{eq:Spincorrelation}
c^{\alpha\alpha^\prime} (\boldsymbol\ell)= \tfrac{1}{3} \langle\, \bf{S}^{\alpha}_{\bf{r}_{0}} \cdot \, \bf{S}^{\alpha^\prime}_{\bf{r}_{0} + \boldsymbol\ell} \,\rangle \, ,
\end{equation}
with $\bf{r}_{0}$ being the position of a given unit cell, while $\alpha$ and $\alpha^\prime$ denote the orbitals $a$, $b$, or $c$. 
The Fourier transform of $c^{\alpha\alpha^\prime}(\boldsymbol\ell)$ is the magnetic structure factor,
\begin{equation}\label{eq:StructFactor}
S(\bf{q}) = \frac{1}{N_{s}} \sum_{\alpha, \alpha^\prime}\sum_{\boldsymbol\ell}c^{\alpha\alpha^\prime}(\boldsymbol\ell)\,e^{\rm{i}\,\bf{q}\cdot\boldsymbol\ell} \, ,
\end{equation}
where $N_s=3L^2$ is the number of sites. It is also instructive to discuss the uniform magnetic susceptibility, which is simply related to the structure factor, Eq.\,\eqref{eq:StructFactor}, through
\begin{equation}
    \chi \equiv \beta S(0,0).
    \label{eq:chi}
\end{equation}
Similarly, one may probe the metallic or insulating features by the electronic compressibility,
\begin{equation}
    \kappa = \frac{1}{\rho^2} \frac{\partial \rho}{\partial \mu},
    \label{eq:Compress}
\end{equation}
where $\rho$ is the global electronic density.

\section{Results}
\label{sec:results}
\subsection{The noninteracting limit ($U=0$)}

\begin{figure}[t] 
	\includegraphics[scale=0.55]{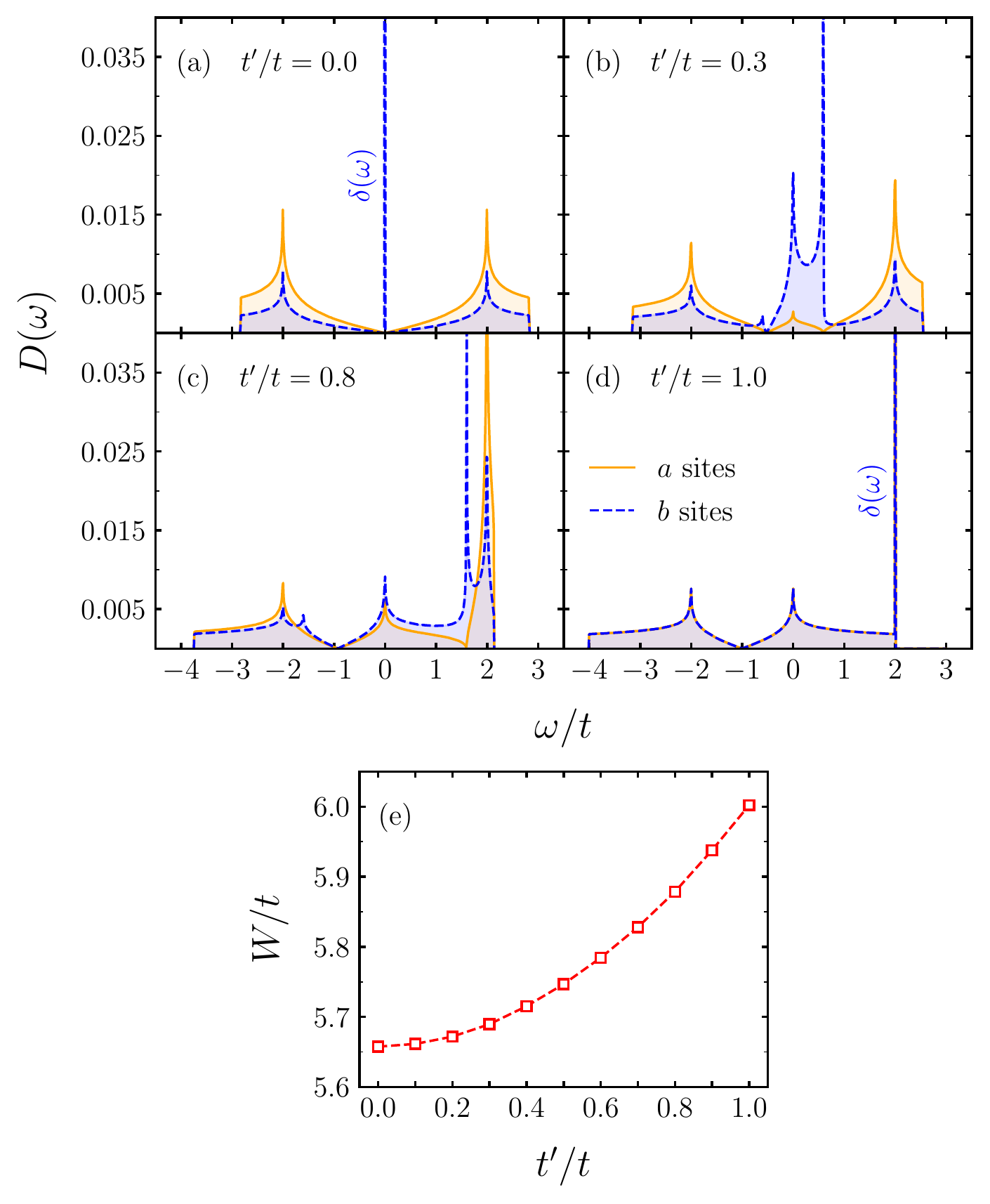}
	\caption{\label{fig:dos-bandwidth} 
 (a)-(d): Site-resolved non-interacting DOS for tight-binding fermions on a kagome lattice for different values of $t^\prime/t$, the ratio between hoppings along $bc$ sites and along sites involving $a$ sites (see Fig.\,\ref{fig:lattice}); $\hbar\omega$ is measured relative to the Fermi energy for half filling.
 (e): Bandwidth, $W/t$, as a function of $t^\prime/t$; the dashed line is a guide to the eye. 
 }
\end{figure}

In the absence of interactions, and setting $t^\prime=0$, i.e.~in the Lieb lattice case, the diagonalization of $\widehat{\mathcal{H}}_\rm{K}$, Eq.\,(\ref{eq:Kinetic_energy}), is straightforward in $\rb{k}$-space.
It yields two dispersive bands, 
\begin{equation}
\epsilon_\pm(\rb{k})=\pm 2t \sqrt{\cos^2 (k_x/2) + \cos^2(k_y/2)},    
\end{equation} 
and a third dispersioneless (flat) one, $\epsilon(\rb{k})=0$,
associated with the $b$ and $c$ sites.
This gives rise to the particle-hole symmetric density of states (DOS) shown in Fig.\,\ref{fig:dos-bandwidth}(a).
Similarly, in the Kagome limit, $t^\prime=1$, the diagonalization yields two dispersive bands,
\begin{equation}
\epsilon_\pm(\rb{k})=-t [1 \pm \sqrt{4f(\rb{k}) - 3}\,],
\end{equation}
with $f(\rb{k}) = \cos^2(k_x/2) + \cos^2(k_y/2) + \cos^2((k_x-k_y)/2)$, and one dispersionless band,
$\epsilon(\rb{k})=2t$.
Figure \ref{fig:dos-bandwidth}(d) shows the corresponding DOS, with the flat band located at the top of the spectrum.

For $0<t^\prime/t<1$, the diagonalization of the non-interacting Hamiltonian is carried out numerically, and leads to the DOS displayed in Fig.\,\ref{fig:dos-bandwidth}(b)-(c).
These panels show that as $t^\prime$ increases the flat band widens, giving way to two van Hove singularities (vHS): one remains at $\omega=0$, while the other moves towards the upper band edge, finally merging with another vHS to form another flat band at $\omega=2t$ when $t^\prime=t$.
Further, as shown in Fig.\,\ref{fig:dos-bandwidth}(e), the bandwidth increases monotonically as $t^\prime$ increases; the relatively small increase may be attributed to the fact that the kinetic energy is insensitive to hopping anisotropy, and therefore significant changes in total bandwidth are essentially driven by the $U/t$ ratio. 
Indeed, with increasing $U/t$ even at small values, the total spectral density is significantly affected, increasing the total bandwidth \cite{Imai03}.

%
\begin{figure}[!t] 
	\includegraphics[width=\linewidth]{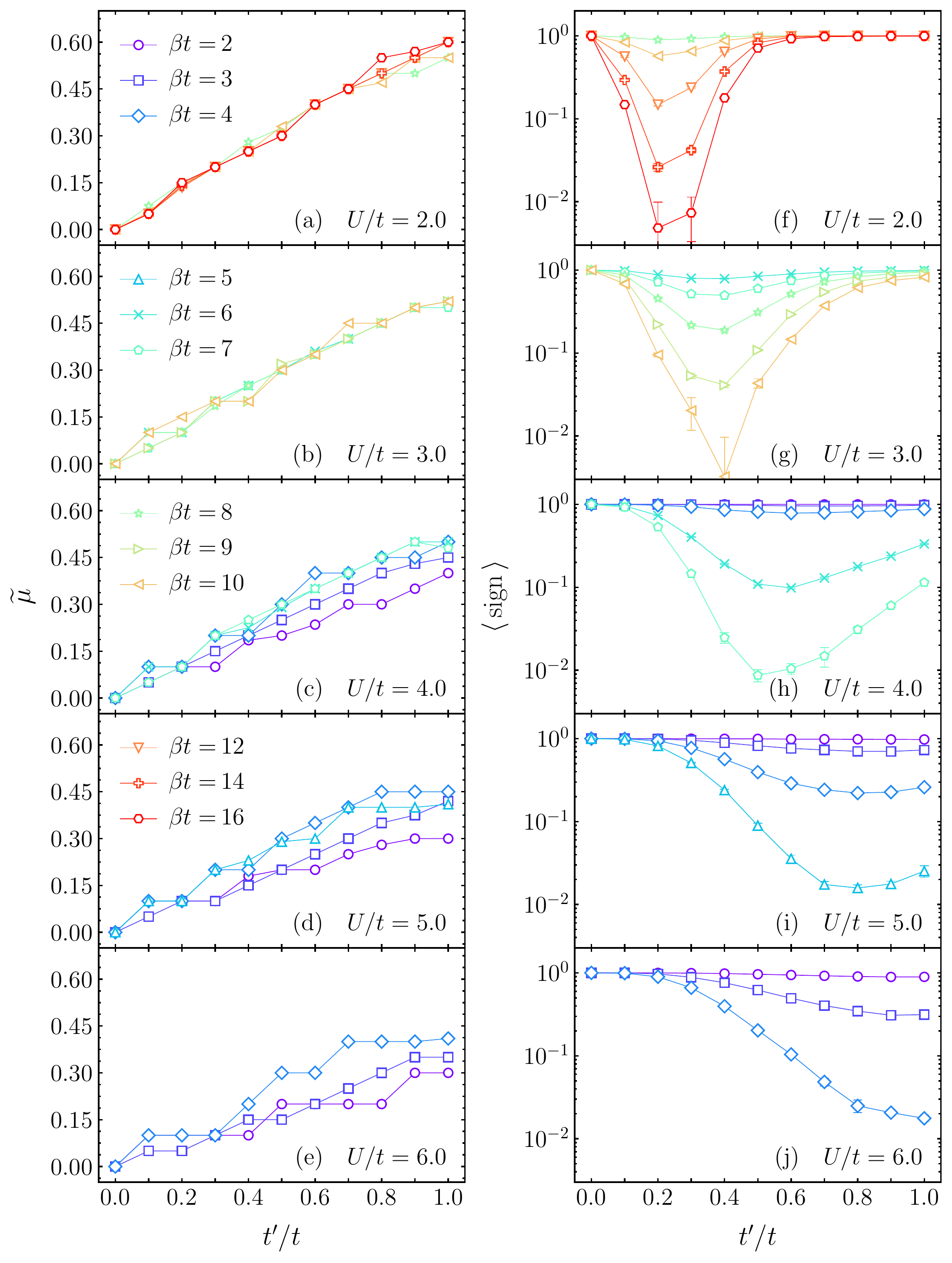}
	\caption{\label{fig:ChemicalPotential_sign} 
    Left panels (a)-(e): The chemical potential, $\widetilde{\mu}$, leading to a half filled band as a function of hopping anisotropy. 
    Each of the panels corresponds to a fixed value of $U/t$, and, in every panel, the curves are for different inverse temperatures, $\beta t$, as indicated in (a)-(b). 
    Right panels (f)-(j): Average sign of the fermionic determinant in a log-linear scale. The system parameters are in strict correspondence with those for the (a)-(e) panels. 
    All data are for a linear lattice size $L=6$, hence with $N_s=3L^2$ sites.
    }
\end{figure}

\subsection{The chemical potential and $\ave{\rm{sign}}$}
\label{subsec:chempot}

The Hamiltonian, Eq.\,(\ref{eq:Hamiltonian}), only displays PHS when $t^\prime=0$, in which case half filling, $\rho=1$, occurs exactly at $\mu=0$.
When $t^\prime\neq0$ the chemical potential leading to  $\rho=1$, $\widetilde{\mu}$, depends on $U$, $t^\prime$, temperature, and, to a lesser extent, on the linear system size, $L$.  
The panels on the left hand side of Fig.\,\ref{fig:ChemicalPotential_sign} show $\widetilde{\mu}$ as a function of hopping anisotropy, for different interaction strengths, $U/t$, and temperatures.
We see that for a given temperature, $\widetilde{\mu}/t$ varies  between 0 and typically 0.5-0.6, as $t^\prime/t$ goes from 0 to 1; further, the larger the repulsion, the stronger is the dependence of  $\widetilde{\mu}$ with the temperature. 
The panels on the right hand side of Fig.\,\ref{fig:ChemicalPotential_sign} show $\ave{\text{sign}}$ as a function of hopping anisotropy, for different interaction strengths, $U$, and temperatures. 
As expected, as the temperature is lowered for a given $U$, the dips in $\ave{\text{sign}}$ become deeper, and, further, as the on-site repulsion increases, the dips move towards the isotropic region while they also widen; see Fig.\,\ref{fig:ChemicalPotential_sign}(f)-(j). 
These panels therefore map out the regions where a much larger number of Monte Carlo sweeps (on the order of $10^6$) is required to mitigate the minus-sign problem, so that we concentrate on parameter ranges leading to $\ave{\text{sign}}\gtrsim 0.1$.

\begin{figure}[t] 
    \includegraphics[width=\linewidth]{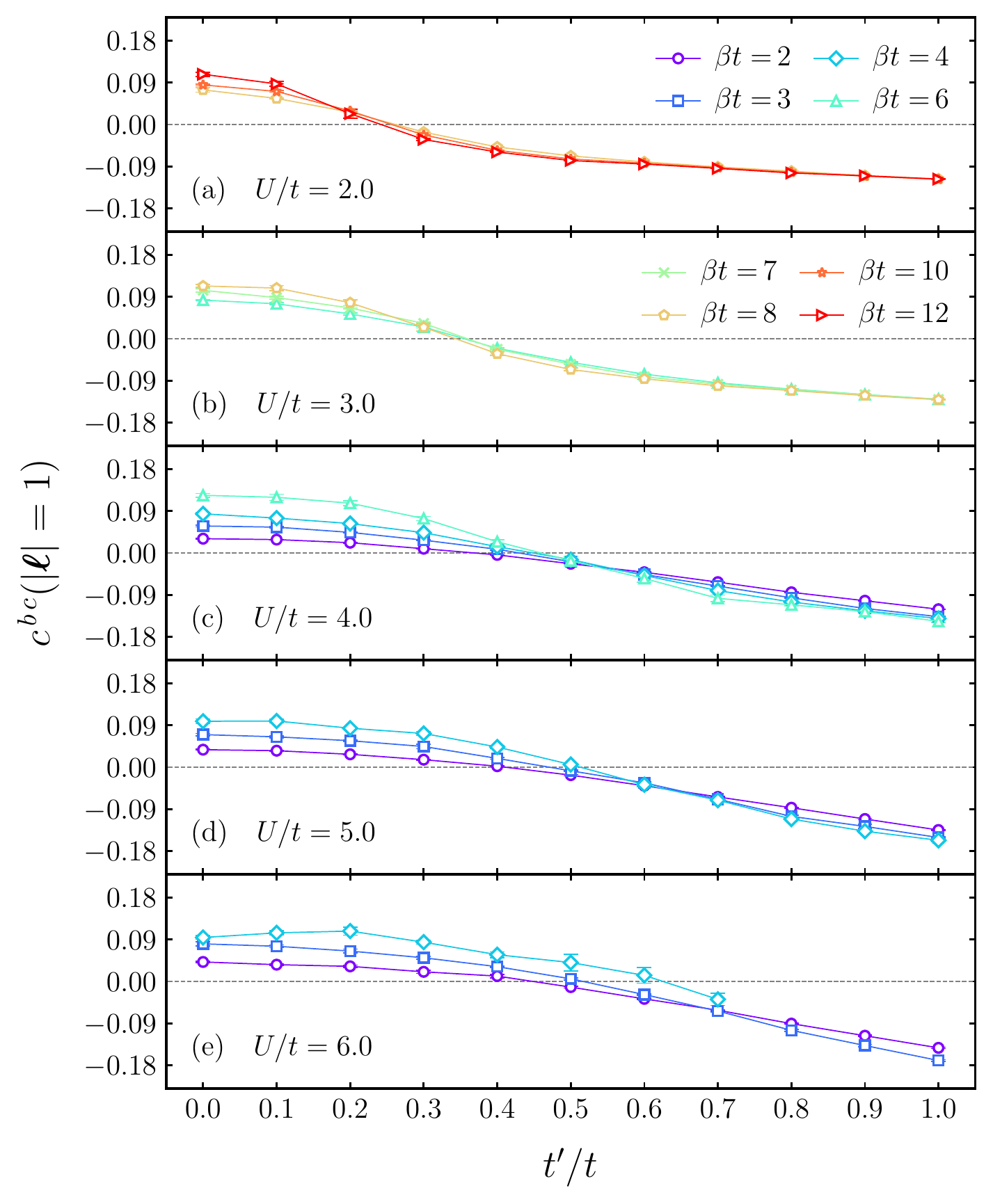}
    \caption{\label{fig:SpinCorrelationNN_bc} 
    Correlations between spins on nearest sites $b$ and $c$, as functions of $t^\prime/t$ at different temperatures, and different values of $U/t$. Each panel corresponds to a given value of $U/t$, and the linear system size is $L=6$. When not shown, error bars are smaller than the symbol sizes.
    }
\end{figure}

\subsection{Magnetic properties}
\label{subsec:MagTransp}

Figure \ref{fig:SpinCorrelationNN_bc} displays the behavior of the spin-spin correlations on nearest $b$ and $c$ orbitals, as $t^\prime/t$ increases from 0 to 1, for different interaction strengths, $U/t$. 
Each panel shows that the correlations are maximally positive in the LL limit, and decrease monotonically towards negative values as $t^\prime/t$ increases towards the KL limit. 
We can also see that for a given $t^\prime/t$, these correlations increase in magnitude as the temperature decreases, a manifestation of their robustness.
Thus, the picture that emerges is that a long-ranged ferrimagnetic state \cite{shen94} at $t^\prime/t=0$ evolves towards a state with strong short-ranged antiferromagnetic correlations in the KL limit as $t^\prime/t$ increases \cite{Furukawa10,andressa22}.

Given this, we note that the sign of $c^{\alpha\alpha^\prime}(\boldsymbol\ell)$ changes at values of $t^\prime/t$ which increase as $U/t$ increases; these values provide a rough estimate for the boundary of the ferrimagnetic phase, $t^\prime_c(U/t)$, as shown in Fig.\,\ref{fig:phases_DQMC}. 
The growth of $t^\prime_c/t$ with $U/t$ can be understood by the fact that the ferrimagnetic state on the LL may be thought of as due to the formation of triplets in the $b$ and $c$ orbitals on the same sublattice, which is more robust the stronger the on-site repulsion is \cite{Costa16}.

\begin{figure}[t] 
    \includegraphics[width=\linewidth]{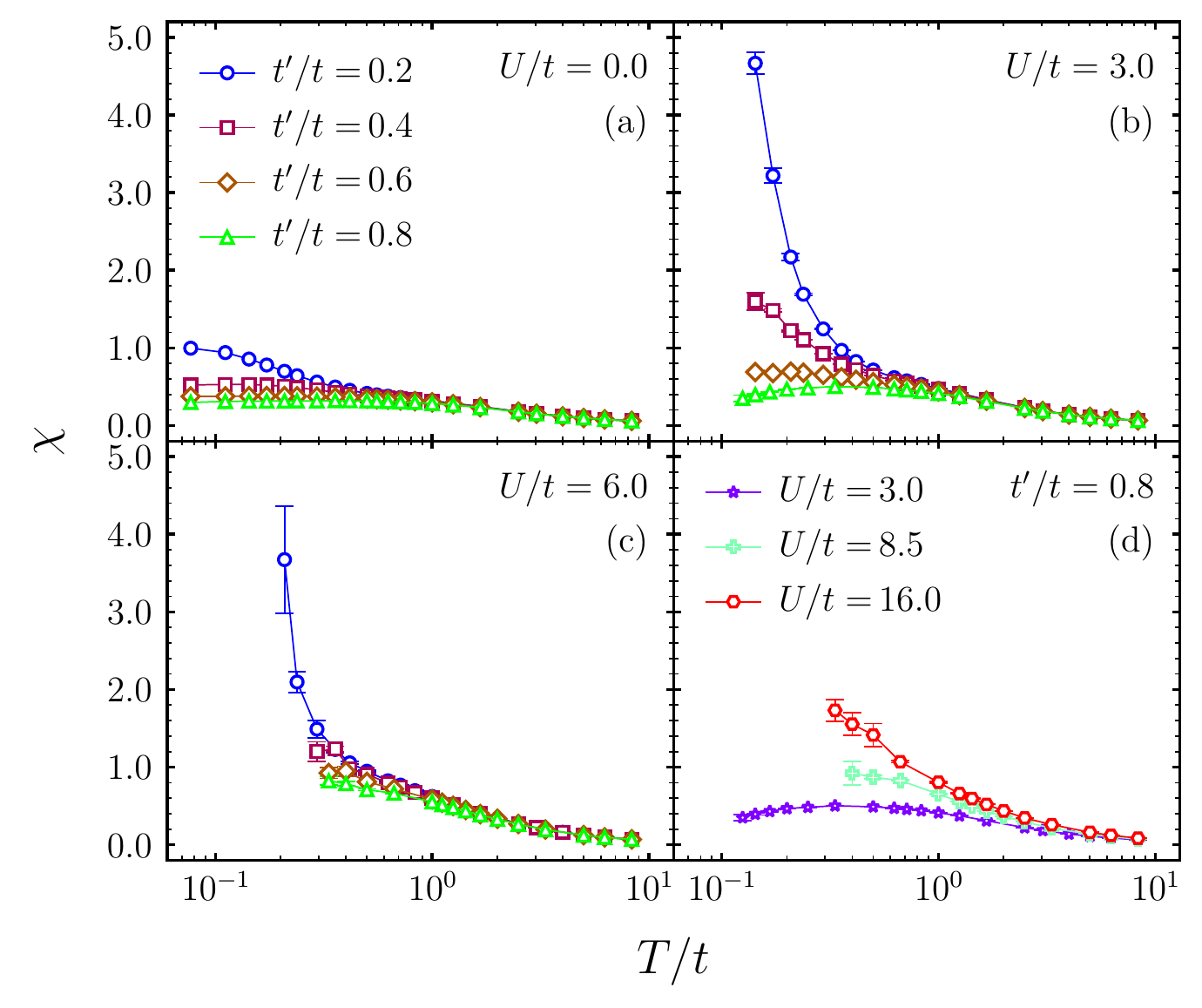}
    \caption{
	Temperature dependence of the uniform susceptibility in linear-log plots. Each of the panels (a) to (c) shows data for fixed $U/t$ and different $t^\prime/t$; panel (d) shows data for fixed $t^\prime/t=0.8$, and different values of $U/t$. 
    Data have been obtained for $L=6$, except in (a), in which case $L=18$ was used.
    When not shown, error bars are smaller than the data points.
    }
    \label{fig:Homo_susceptibility}
\end{figure}

These analyses are supported by the temperature dependence of the uniform susceptibility, Eq.\,\eqref{eq:chi}.
In the absence of interactions, Pauli behavior is verified for all $t^\prime/t$; see Fig.\,\ref{fig:Homo_susceptibility}(a).
For $U/t=3$, different values of $t^\prime/t$ cause different responses, as shown in Fig.\,\ref{fig:Homo_susceptibility}(b): 
for $t^\prime/t=0.2$ $\chi$ shows a Curie-like behavior indicating a ferrimagnetic ground state, while for $t^\prime/t\geq 0.6$ Pauli behavior sets in. 
The behavior for $t^\prime/t=0.4$ is borderline, being consistent with the change in sign of the correlation functions shown in Fig.\,\ref{fig:SpinCorrelationNN_bc}(b).
For $U/t=6$, Fig.\,\ref{fig:Homo_susceptibility}(c) shows that the Curie-like behavior is still present for $t^\prime/t=0.2$, but the minus-sign problem prevents us from decreasing the temperature  
below $T/t\approx 0.3$ for $t^\prime/t\geq 0.4$.
Nonetheless, according to Fig.\,\ref{fig:SpinCorrelationNN_bc}(e), for $t^\prime/t < 0.6$ the system still displays ferrimagnetic correlations, so that a rise of $\chi$ should not be discarded if the temperature is lowered.
And, finally, data for $\chi(T)$ in Fig.\,\ref{fig:Homo_susceptibility}(d) correspond to fixed $t^\prime/t=0.8$, and different on-site interactions.
While for $U/t=3$ and $U/t=8.5$ the behavior is Pauli-like, for $U/t=16$ one detects a tendency to increase as $T/t$ is lowered, similarly to what was found in the KL limit \cite{andressa22}.

\begin{figure}[t] 
	\includegraphics[width=\linewidth]{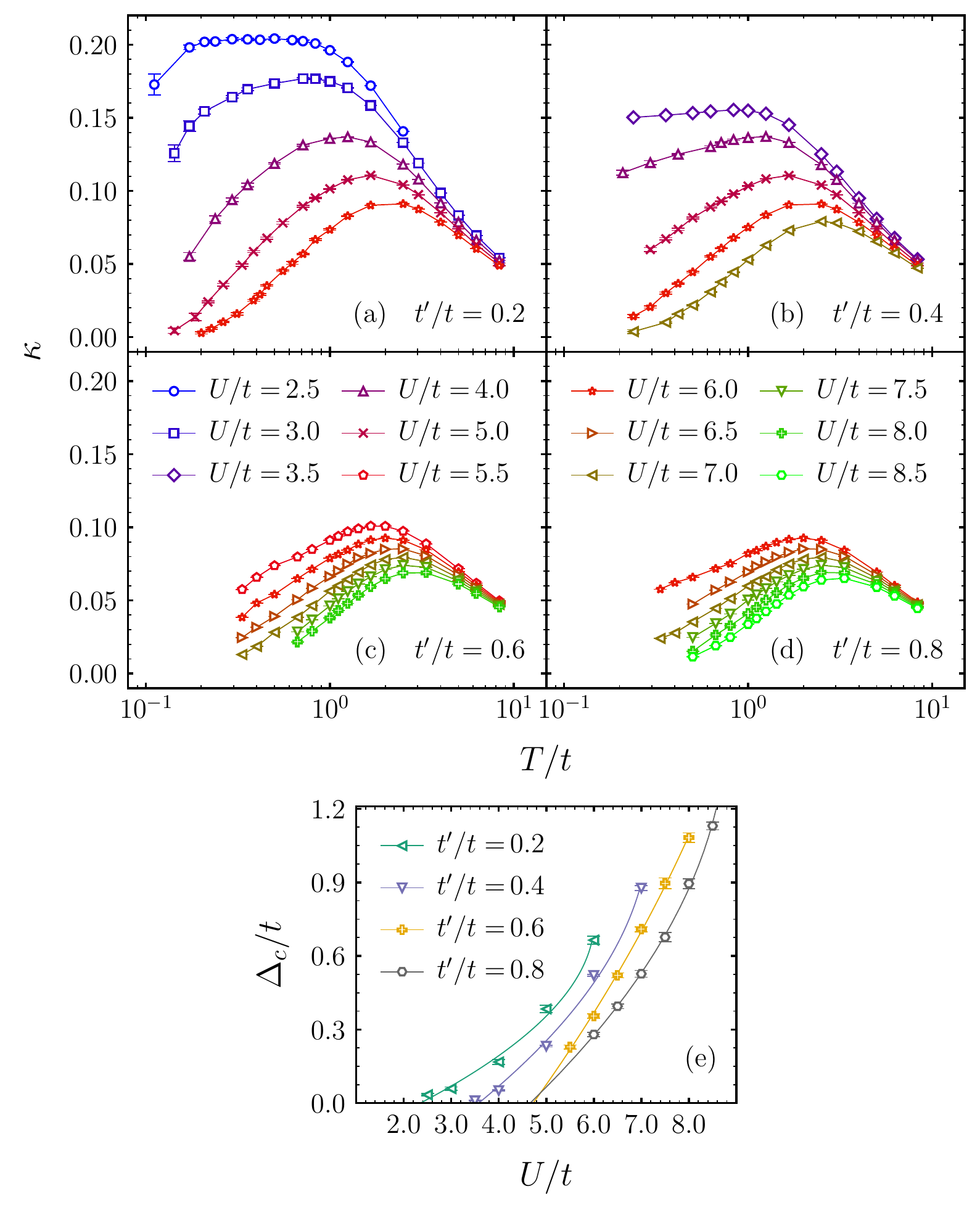}
	\caption{ 
    Panels (a)-(d): Compressibility as a function of temperature, for fixed values of $t^\prime/t$, and different values of $U/t$. 
    (e) Estimates of the gap obtained by fitting an exponential form, $\kappa\propto\exp{(-\Delta_c/k_\rm{B}T)}$, to the low temperature data of panels (a)-(d); see text. When not shown, error bars are smaller than the symbol size, while the lines are parabolic fits to the curves. 
    }
 \label{fig:Compress_and_gap}
\end{figure}
\subsection{Transport properties}
\label{subsec:Transport}

Now we proceed by examining  transport properties, namely the electronic compressibility, Eq.\,\eqref{eq:Compress}. Interaction-driven metal-insulator transitions are usually signalled by the opening of a single-particle gap at the Fermi energy; this gap appears as a plateau in plots of $\rho(\mu)$ at fixed low temperatures, and, in turn, leads to an exponential decay of $\kappa$ at low temperatures,  $\kappa\propto\exp{(-\Delta_c/k_\text{B}T)}$ -- here we define $k_\rm{B} \equiv 1$.
This behavior is observed in Figs.\,\ref{fig:Compress_and_gap}(a)-(d), in particular for large values of $U/t$.

By fitting the low temperature data of Figs.\,\ref{fig:Compress_and_gap}\,(a)-(d), we obtain the dependence of $\Delta_c/t$ with $U/t$ shown in Fig.\,\ref{fig:Compress_and_gap}\,(e). 
The values of $U/t$ at which $\Delta_c/t \to 0$ provide estimates for $U_c^\text{M}/t$, the critical values of $U/t$ above which the system is an insulator.

\subsection{The Phase Diagram}
\label{subsec:phasediag}

\begin{figure}[t] 
    \includegraphics[width=\linewidth]{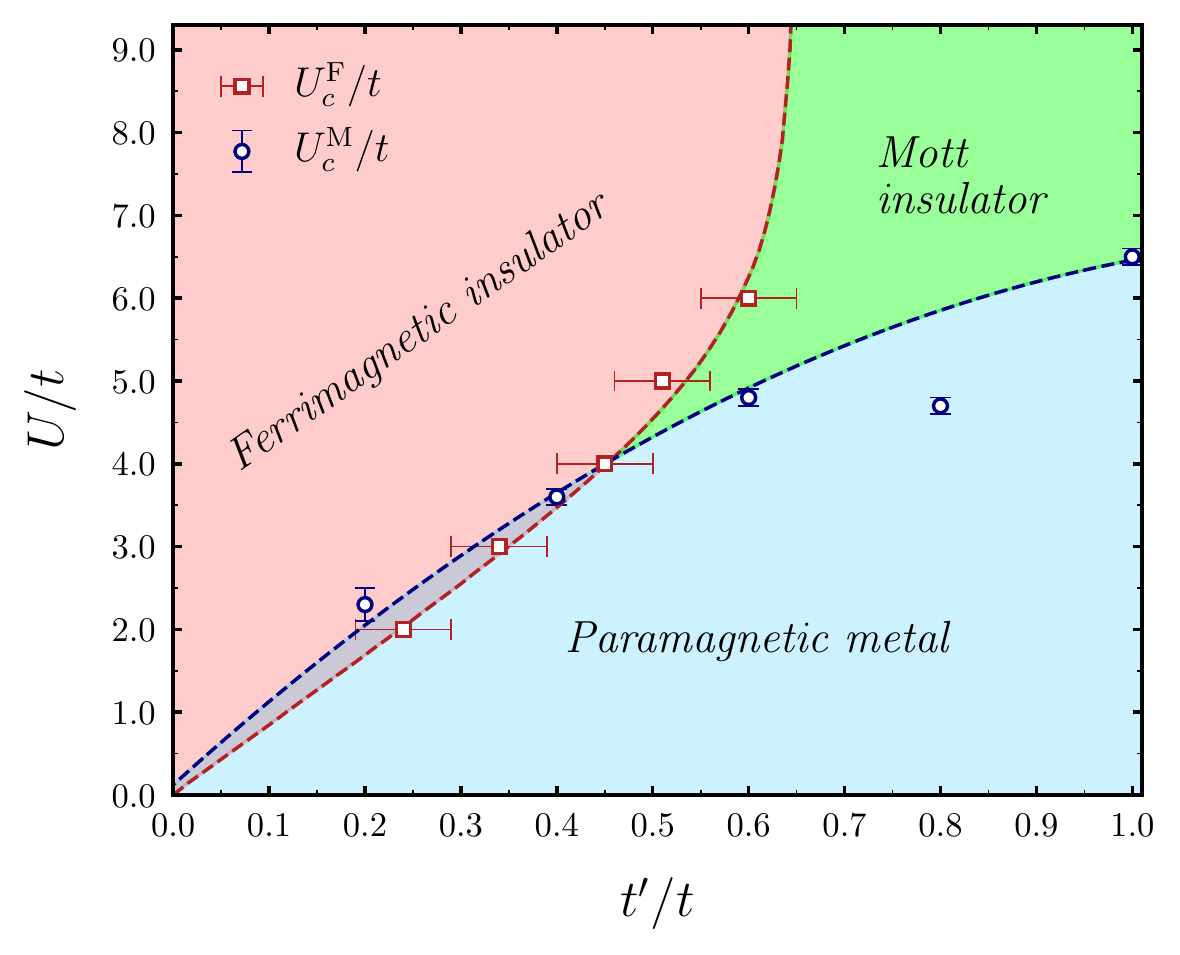}
	\caption{\label{fig:phases_DQMC} 
 Phase diagram of the Hubbard model on the anisotropic kagome lattice at half-filling. 
 Critical points labelled $U_c^\text{F}/t$ and $U_c^\text{M}/t$ have been determined through data in Fig.\,\ref{fig:SpinCorrelationNN_bc} and Fig.\,\ref{fig:Compress_and_gap}(e), respectively; see text.
 The dashed lines through the critical points are guides to the eye.
 }
\end{figure}

The analyses of the preceding subsections may be summarized in the phase diagram shown in Fig.\,\ref{fig:phases_DQMC}.
The change in sign of correlation functions provides an estimate for $t_c^\prime/t(U/t)$ for a given temperature; see Fig.\,\ref{fig:SpinCorrelationNN_bc}. 
Since the minus sign prevents us from obtaining a significative sequence of data at different temperatures for all values of $U/t$, we adopt as $t_c^\prime/t(U/t)$ the value at which the curve for the lowest temperature crosses the horizontal axis. 
The error in this estimate is provided by the grid of $t^\prime/t$ values. 
With this we are able to set up a boundary for the ferrimagnetic state, whose points are denoted by $U_c^\text{F}/t$ in Fig.\,\ref{fig:phases_DQMC}.

The boundary of the metallic region, on the other hand, is obtained through the analysis of the single-particle gap. 
As mentioned in Sec.\,\ref{subsec:Transport}, for a fixed $t^\prime/t$, we determine $U_c^\text{M}/t$ as the value at which $\Delta_c\to0$. 
The errors, in this case, are those emerging from parabolic fits of $U_c/t(\Delta_c/t)$ for the curves in Fig.\,\ref{fig:Compress_and_gap}\,(e).
Interestingly, both ferrimagnetic insulator and paramagnetic metal curves are very close to each other for small values of $t^\prime/t$, suggesting that both transitions occur simultaneously.
However, from Fig.\,\ref{fig:phases_DQMC}, we see that such behavior changes for larger frustration, with a bifurcation at $t^\prime/t\approx 0.45$ with a slight increase of $U_c^\text{F}/t$ up to $t^\prime/t\approx 0.6$ -- beyond this point, one cannot find ferrimagnetism.
Determining the features of the region that is neither metallic nor ferrimagnetic is challenging.
Due to its similarities with the Mott region of the Kagome lattice, $t^{\prime}/t = 1$, in particular the strong short-range antiferromagnetic correlations, we define it as a Mott insulator region, and we expect that spin liquid features may occur for sufficiently large $U/t$. 

It is important to note that our data for $\ave{\text{sign}}$ verify the relation with quantum critical points \cite{Mondaini22,mondaini2022_2}. 
Starting with $U/t=2$, Fig.\,\ref{fig:ChemicalPotential_sign} shows that the dip in $\ave{\text{sign}}$ for $\beta t =16$ occurs near $t_c^\prime/t=0.20\pm0.05$. 
Similarly, for $U/t=3$ the dip for $\beta t=10$ occurs at $t_c^\prime/t=0.40\pm 0.05$, which is also in accord with the critical point indicated in Fig.\,\ref{fig:phases_DQMC}.
For $U/t=4$, the minimum of $\ave{\text{sign}}$ moves to $t_c^\prime/t=0.50\pm 0.05$. 
This trajectory of $t_c^\prime/t$ therefore provides numerical evidence that $\ave{\text{sign}}$ is picking up the transitions labelled $U_c^\text{M}/t$ in Fig.\,\ref{fig:phases_DQMC}.
Indeed, although the minima of $\ave{\text{sign}}$ broaden considerably for $U/t\gtrsim 5$, they are consistent with a smaller rate of change in $U_c^\text{M}/t$ in this region.
Interestingly, one cannot ascertain whether the dominant cause of a drop in $\ave{\text{sign}}$ is the transition between a paramagnet and a magnetically ordered state (irrespective of being ferrimagnetic or antiferromagnetic) or between a metal and an insulator.

As mentioned before, this model has been studied within a mean-field (MF) approach \cite{Yamada11}. 
Although restricted to the range $0.5 \leq t^\prime/t \leq 1$, the phase diagram thus obtained identifies a metal-insulator (MI) transition, in addition to the ferrimagnetic-Mott transition in the insulating regime. 
Nonetheless, some quantitave differences are worth mentioning:
First, the Paramagnetic metal - MI phase boundary obtained through the MF approximation displays a slower increase with $t^\prime/t$, than the one obtained here. Second, the steep FM-Mott Insulator boundary lies around  $t^\prime/t=0.8$  in \cite{Yamada11}   larger than our estimate, of around  $t^\prime/t = 0.6$.


\section{Conclusions}
\label{sec:conc}

We have examined the Hubbard model on a Kagome lattice with hopping anisotropy such that it allows us to continuosly interpolate between the unfrustrated Lieb lattice and the fully frustrated Kagome lattice.
This interpolation is achieved by varying the hopping amplitude, $0\leq t^\prime/t\leq1$, along one of the diagonals of the Kagome lattice.
Through extensive quantum Monte Carlo simulations we have calculated different magnetic and transport responses, from which we have obtained a phase diagram in the $(U/t,t^\prime/t)$ parameter space. 
We have also analyzed the average sign of the fermionic determinant, and found that it provides consistent predictions for critical points, as recently proposed \cite{Mondaini22,mondaini2022_2}.  

The picture that emerges is that a metal-insulator transition takes place at some $U_c^\text{M}/t (t^\prime/t)$, which increases monotonically with $t^\prime/t$, from $U_c^\text{M}/t(0)=0$. 
We have also found that increasing this type of frustration causes a phase transition between a ferrimagnetic phase and a Mott phase. 
We hope our findings will motivate further studies of ultracold fermionic atoms on Lieb and Kagome lattices.

\section*{Acknowledgments}
The authors are grateful to the Brazilian Agencies Conselho Nacional de Desenvolvimento Cient\'\i fico e
Tecnol\'ogico (CNPq), Coordena\c c\~ao de Aperfei\c coamento de Pessoal de Ensino Superior (CAPES),
 and Instituto 
Nacional de Ciência e Tecnologia de Informação Qu\^antica (INCT-IQ) for funding this project.
Financial support from CNPq, Grant No. 313065/2021-7 (N.C.C.), and Grant Nos 308335/2019-8 and 403130/2021-2 (T.P.)  and from FAPERJ -- Funda\c{c}\~ao Carlos Chagas Filho de Amparo \`a Pesquisa do Estado do Rio de Janeiro --, Grant No.\,200.258/2023 (SEI-260003/000623/2023) (N.C.C) and Grant numbers E-26/200.959/2022 and  
E-26/210.100/2023 (T.P.).


\bibliography{lima_kagome.bib}
\end{document}